\documentclass[letter,twocolumn]{jpsj2}

\title{Magnetic Domain Patterns Depending on the Sweeping Rate of
Magnetic Fields}

\author{%
Kazue \textsc{Kudo}\thanks{E-mail address: 
kudo@a-phys.eng.osaka-cu.ac.jp}, 
Michinobu \textsc{Mino}$^{1}$ and Katsuhiro \textsc{Nakamura}
}

\inst{Department of Applied Physics, Graduate School of Engineering, 
Osaka City University, Osaka 558-8585 \\
$^{1}$Department of Physics, Graduate School of Natural Science and
Technology, Okayama University, Okayama 700-8530}

\abst{%
The domain patterns in a thin ferromagnetic film are investigated in both
experiments and numerical simulations.
Magnetic domain patterns under a zero field are usually observed after an
external magnetic field is removed. 
It is demonstrated that 
the characteristics of the domain patterns depend on the decreasing rate of
the external field, although it can also depend on other factors.
Our numerical simulations and experiments show the following properties
of domain patterns:
a sea-island structure appears when the field decreases rapidly from the 
saturating field to the zero field, while
a labyrinth structure is observed for a slowly decreasing field.
The mechanism of the dependence on the field sweeping rate is discussed
in terms of the concepts of crystallization.
}

\kword{domain pattern, pattern formation, garnet thin film, domain
dynamics, time-dependent field}
\begin{document}
\maketitle

There are a large number of physical and chemical systems that display
domain patterns.\cite{seul} 
Ferromagnetic garnet thin films also show domain patterns, which can have
various kinds of structures. For example, when an external field is
applied, a hexagonal lattice, bubbles, or stripes are observed. Under a
zero field, usually 
a labyrinth structure is observed, although a parallel-stripe structure is
more stable than a labyrinth structure.
In fact, it is observed in experiments that a labyrinth structure
changes to a parallel-stripe structure under field
cycles.\cite{miura,mino}
However, the domain patterns observed under a zero field do not always have  
a labyrinth or parallel-stripe structure. 
In some samples, under a certain condition, 
a sea-island structure is observed, which consists of many small up-spin
(down-spin) domains surrounded by a ``sea'' of down (up) spins.  

In order to observe the magnetic domain patterns under a zero field in
experiments, usually
an external magnetic field is applied once up to the saturation field,
and then it is removed. 
Here, we will focus on the domain patterns under a zero field and their
dependence on the decreasing rate of the external field, although
it is natural that the characteristics of the observed patterns can depend
on the properties of the samples, \textit{e.g.}, disorder, anisotropy,
interactions among spins, and so on. 
In fact, there have been several simulations that show how domain
patterns 
change depending on such properties under a very slowly changing
field.\cite{jagla04,jagla05,deutsch}
There also exist some works on the dependence of
hysteresis on the frequency and amplitude of an oscillating
field.\cite{luse,sides,jang}  
However, for the characteristics of domain patterns under a zero field,
their dependence on the changing rate of a 
time-dependent field has not been investigated so far.

In this letter, we report both the experimental observations and the
numerical simulations of domain patterns. In particular, we show how the 
characteristics of domain patterns depend on the sweeping rate of the
magnetic field. 
First, the experimental procedure and results are explained. 
Next, the model of the
system and numerical procedure are described. After displaying numerically
simulated domain patterns, we study the field
dependence of the characteristics of domain patterns, i.e., the number of
domains and the domain area.
The mechanism of the dependence will also be discussed. 

\begin{figure}[tb]
\begin{center}
\includegraphics[width=8cm]{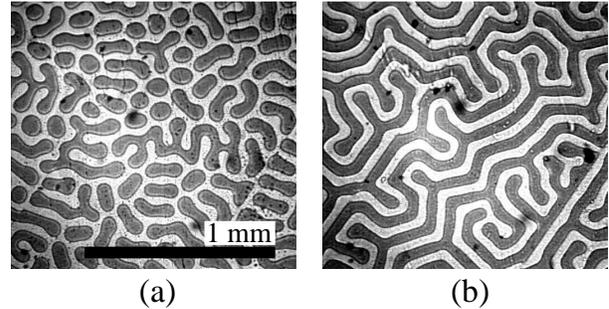}
\end{center}
\caption{Experimentally observed domain patterns under a zero field.  
The external field was decreased at a (a) fast rate and (b) slow
 rate. (See the text.)}
\label{fig:EX}
\end{figure}

Experiments were performed on ferrimagnetic garnet films of composition
(BiGdY)$_3$(FeGa)$_5$O$_{12}$ grown epitaxially on the (111) face
of gadolinium gallium garnet substrates.
This film has strong uniaxial
magnetic anisotropy with the easy axis perpendicular to the film. The
perpendicular saturating field is about 120 Oe. Magnetic domain
structures are easily observed by using a polarizing microscope with the
help of the magneto-optical Faraday effect. 
Figure~\ref{fig:EX}(a) shows the domain
pattern under a zero field when the magnetic field has been decreased from
220 to 0 Oe at a fast sweeping rate of $2\times10^5$ Oe/s.  
This picture shows a
sea-island structure. The observed structure at a sweeping rate of 10
Oe/s is 
shown in Fig.~\ref{fig:EX}(b). At this slow rate, the domain pattern has
a labyrinth structure. 

\begin{fullfigure}[tb]
\begin{center}
\includegraphics[width=15cm]{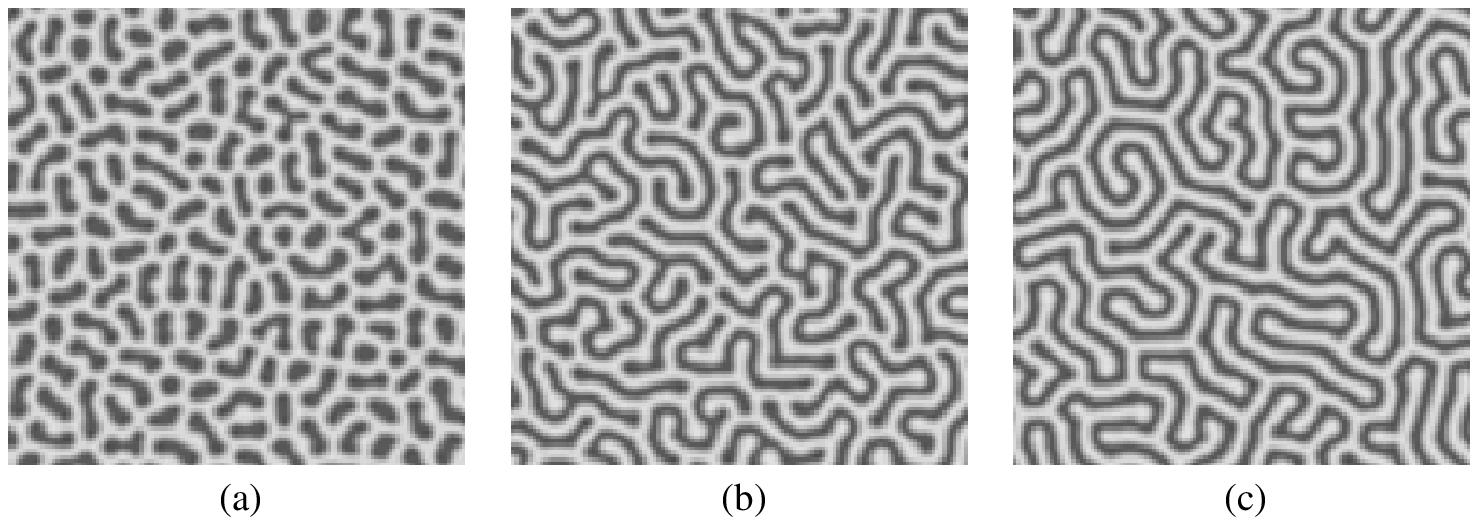}
\end{center}
\caption{Numerically simulated domain patterns under a zero field
 ($1/4$-parts of the whole system). 
 The value of $\phi(\mib{r})$ is positive and negative in the white
 and black areas, respectively. The decreasing rate of the external field
 is (a) $v=10^{-2}$, (b) $v=10^{-3}$, and (c) $v=10^{-4}$.}
\label{fig:PT}
\end{fullfigure}
In our numerical simulations, we apply a simple two-dimensional
Ising-like model (see refs.~\citen{jagla04}, \citen{jagla05}, and
references therein). 
We consider a scalar spin field $\phi(\mib{r})$, where
$\mib{r}=(x,y)$. Namely, the model is a continuous model.
The positive and negative values of $\phi(\mib{r})$ correspond to the up
and down spins, respectively.
Our model Hamiltonian consists of four energy terms:
uniaxial-anisotropy energy $H_{\rm ani}$, local ferromagnetic interactions
(exchange interactions) $H_J$, long range dipolar interactions 
$H_{\rm di}$, and interactions with the external field $H_{\rm ex}$.
The anisotropy energy term implies that $\phi(\mib{r})$ tends to take
either one of 
two different values [$\phi(\mib{r})=\pm 1$]: 
\begin{equation}
 H_{\rm ani}=\alpha\lambda(\mib{r}) \int {\rm d}\mib{r} \left(
-\frac{\phi(\mib{r})^2}{2}+\frac{\phi(\mib{r})^4}{4}
\right),
\label{eq:Ha}
\end{equation}
where the factor $\lambda(\mib{r})$ expresses disorder, and
\begin{equation}
 \lambda(\mib{r})=1+\mu(\mib{r})/4.
\label{eq:disorder}
\end{equation}
Here, $\mu(\mib{r})$ is an uncorrelated random number with a Gaussian
distribution whose average and variance are 0 and $\mu_0^2$, respectively. 
However, $\mu(\mib{r})$ should have a cutoff so that $\lambda(\mib{r})$
is always 
positive. 
We consider the disorder effect only in the anisotropy term, although
some effects of disorder could also be applied to the other interaction
terms. 
The ferromagnetic (exchange) and dipolar interactions are described by 
\begin{equation}
 H_J=\beta\int {\rm d}\mib{r} \frac{|\nabla\phi(\mib{r})|^2}{2}
\label{eq:Hj}
\end{equation}
and
\begin{equation}
 H_{\rm di}=\gamma\int {\rm d}\mib{r} {\rm d}\mib{r}'
 \phi(\mib{r})\phi(\mib{r}') G(\mib{r},\mib{r}'),
\label{eq:Hdi}
\end{equation}
respectively. Here, $G(\mib{r},\mib{r}')\sim |\mib{r}-\mib{r}'|^{-3}$ at 
long distances. 
The integral in eq.~(\ref{eq:Hdi}) should have a lower cutoff since 
$G(\mib{r},\mib{r}')\to\infty$ at short distances.
The term from the interactions with the external field is given by
\begin{equation}
 H_{\rm ex}=-h(t) \int {\rm d}\mib{r} \phi(\mib{r}),
\label{eq:Hex}
\end{equation}
where $h(t)$ is the time-dependent external field.
From eqs.~(\ref{eq:Ha})--(\ref{eq:Hex}), the dynamical equation of our
model is described by  
\begin{eqnarray}
 \frac{\partial \phi (\mib{r})}{\partial t}&=&
 -\frac{\delta (H_{\rm ani}+H_{J}+H_{\rm di}+{H_{\rm ex}})}
 {\delta \phi (\mib{r})} \nonumber\\
&=& \alpha \lambda(\mib{r}) [\phi(\mib{r})-\phi(\mib{r})^3]
+\beta\nabla^2\phi(\mib{r})  \nonumber\\
&&-\gamma\int {\rm d}\mib{r}' \phi(\mib{r}') G(\mib{r},\mib{r}')
+h(t).
\label{eq:A-C}
\end{eqnarray}
The coefficients of the anisotropy energy and the exchange and dipolar
interactions --- $\alpha$, $\beta$, and $\gamma$ respectively
--- are all positive.

In numerical simulations, it is useful to calculate the time evolution of
eq.~(\ref{eq:A-C}) in Fourier space. 
The equation is rewritten as
\begin{equation}
 \frac{\partial \phi_{\mib k}}{\partial t} 
 =\alpha  [\lambda (\phi-\phi^3)]_{\mib{k}}
-(\beta k^2+\gamma G_{\mib{k}})\phi_{\mib{k}}
+h(t)\delta_{\mib{k},0},
\label{eq:k-eq}
\end{equation} 
where $[\cdot]_{\mib{k}}$ denotes the convolution sum and $G_{\mib{k}}$
is the Fourier transform of $G(\mib{r},0)$. In the calculation below, 
if $G(\mib{r},0)\equiv 1/|\mib{r}|^3$, then
\begin{equation}
 G_{\mib{k}}=a_0-a_1 k,
\label{eq:Gk}
\end{equation}
where $k=|\mib{k}|$ and
\begin{equation}
 a_0=2\pi\int_d^\infty r{\rm d}r G(r), \quad a_1=2\pi.
\label{eq:a0a1}
\end{equation}
Here, $d$ is the cutoff length, which is approximately the thickness of
the film in experiments. We set $d=\pi/2$, namely, $a_0=4$, in our
simulations. In the simulations, 
$512 \times 512$
lattices with periodic boundary conditions are used, and we fix two
parameter values: $\beta=2.0$, $\gamma=2\beta/a_1=2.0/\pi$. 
These parameters
determine the characteristic length of domain patterns.
According to the linear analysis of eq.~(\ref{eq:k-eq}), 
$\phi_{\mib{k}}$ of the modes with $k=1.0$ grows the best at these fixed
values. 
Namely, the characteristic wave number $k_0$ of the domain patterns is
unity: $k_0=1.0$.
Moreover, we fix the other two parameters as follows: $\alpha=2.5$ and
$\mu_0=0.3$. It would be interesting to investigate the dependence on
$\alpha$ or $\mu_0$, which will be studied elsewhere. 
Only the changing rate $v$ of $h(t)$ is the control parameter of our
interest:  
\begin{equation}
 h(t)=h_{\rm ini}-vt,
\label{eq:h}
\end{equation}
where $h_{\rm ini}$ is the saturating field; $h_{\rm ini}\simeq 1.4$ 
in this case. 
The value $h_{\rm ini}$ was estimated
from a simulation under an ascending field. In other words, the domains  
first spontaneously form a pattern under a zero field, 
and then an increasing field is applied. 
At a certain value of $h$, the domains with negative $\phi(\mib{r})$
disappear, and we define this value as $h_{\rm ini}$. 


Figure~\ref{fig:PT} shows the domain patterns under a zero field
that are simulated by the abovementioned method.
The white and black areas correspond to positive and negative
$\phi(\mib{r})$, i.e. up and down spins, respectively.
As the initial condition, $\phi(\mib{r})$ was given by random numbers in
the interval $1.0< \phi(\mib{r}) < 1.1$ at $t=0$. The external field
decreases 
from $h_{\rm ini}$ to $0$, following eq.~(\ref{eq:h}). Once the external
field becomes zero at $t_0=h_{\rm ini}/v$, the field remains zero for 
$t\ge t_0$.   
The change in the domain patterns for $t\ge t_0$
is very little in this case. 
For some other sets of parameters ($\alpha$ and $\mu_0$),
however, the characteristics of the domain patterns change after the field
vanishes.\cite{note1}  

The simulation results have the same behavior as the experimental
ones. Namely, a sea-island structure, which consists of many short
domains, appears under a rapidly decreasing field, and a
relatively small number of domains grow to
form a labyrinth structure under a slowly decreasing field.
In other words, both the experimental and numerical results suggest that 
the field sweeping rate affects the domain size (domain length).

The abovementioned results also mean that a domain pattern forms an unstable
structure under a rapidly decreasing field.
Since the ferromagnetic (exchange) interactions prefer small domain
surfaces (domain walls), a structure composed of a few large domains
is more stable than one composed of many small domains.
In other words, a labyrinth structure is more stable than a sea-island
structure from the viewpoint of energy. 
Before discussing why such an unstable structure appears under a
rapidly decreasing field, we 
consider how domains grow to form a structure under a descending field.

\begin{figure}[tb]
\begin{center}
\includegraphics[width=8cm]{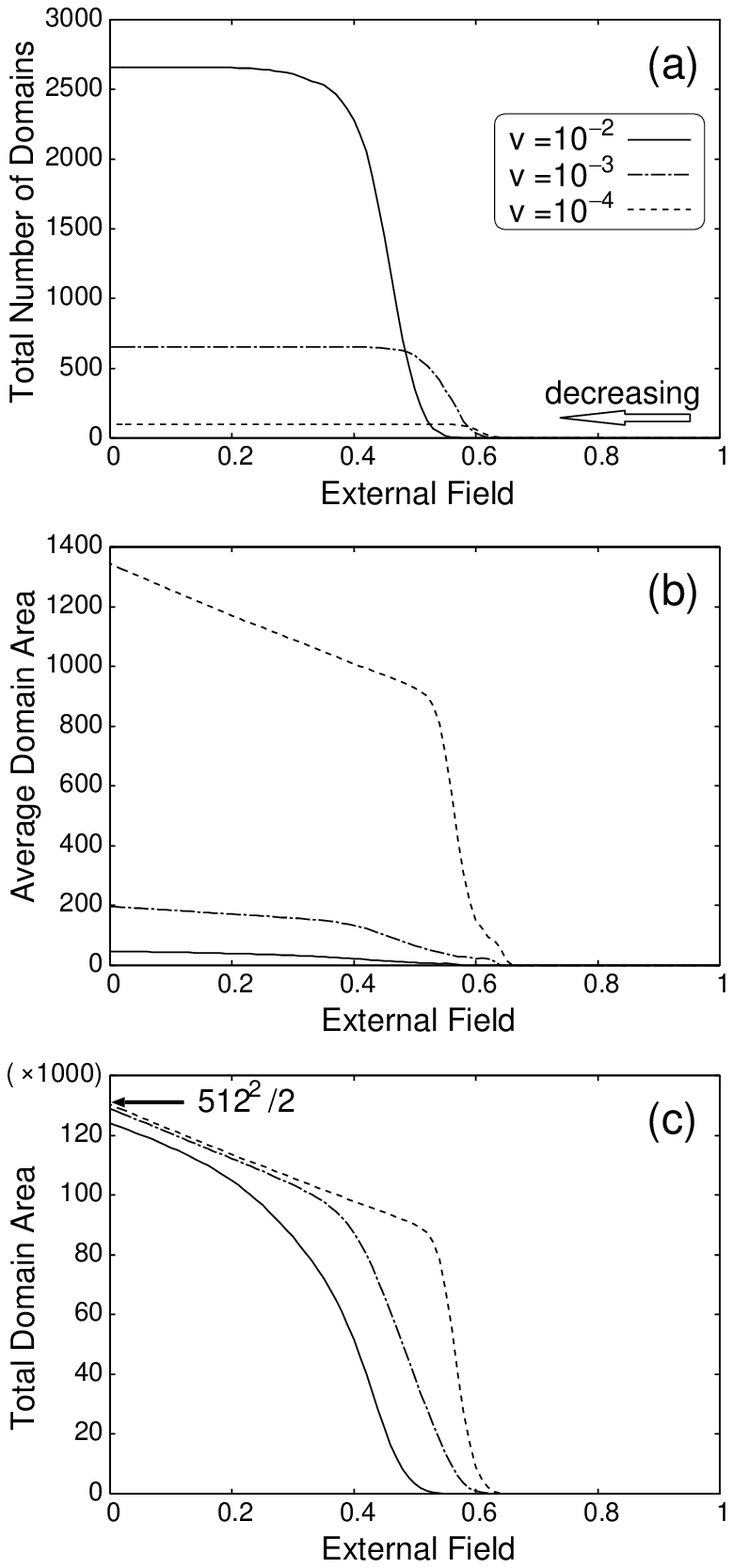}
\end{center}
\caption{The field dependence of (a) total number of domains, (b) average
 domain area, and (c) total domain area. The solid, chained, and dashed
 lines are for $v=10^{-2}$, $v=10^{-3}$, and $v=10^{-4}$, respectively. 
 The domain area is the number
 of grid points that have negative values of $\phi(\mib{r})$. 
 The results under a zero field correspond to Fig.~\ref{fig:PT}.}
\label{fig:stat}
\end{figure}

How are the domain patterns in Fig.~\ref{fig:PT} created? 
The field dependence, which is now identified with the time dependence
due to eq.~(\ref{eq:h}), 
of the number and area of negative-$\phi$ (black) domains is shown in
Fig.~\ref{fig:stat}.   
Under a descending field, the total number of domains for each $v$ rapidly
increases at a certain field and has a plateau in the low-field region.
In the plateau region, each domain grows; however, no new domain appears. 
The length of the plateau is longer for small $v$ than for large $v$.
For the rapidly decreasing field ($v=10^{-2}$), the value of the field
where
the first domain appears is lower than that for the other slowly
decreasing
fields ($v=10^{-3}$, $10^{-4}$). 
From these results, we see that a small number of domains grow for a
long time under a slowly decreasing field and vice versa.

Both the average and total domain areas show rapid growth at
first and slow growth in the low-field region. 
This behavior is explained as follows.
In the rapid growth region, where the domain area increases rapidly,
domains grow freely. In the slow growth region, 
however, domains cannot grow freely since they
tend to repel each other because of the dipolar interactions. Moreover,
domains do not connect with each other for the parameter combination used in
this calculation.\cite{note2}
In the small-$v$ case, since the number of domains is small,
the change in the growth rates becomes clear.
Here, we note that the difference between the total domain area and
$512^2/2$ in Fig.~\ref{fig:stat}(c) corresponds to the magnetization. We
see that there is a small difference between the total domain 
area under a zero field and $512^2/2$, which
corresponds to the remanent magnetization.

Now, let us discuss why the characteristics of domain patterns depend on
the field sweeping rate, as shown in Figs.~\ref{fig:PT} and
\ref{fig:stat}. We can explain the mechanism of this behavior
by using the concept of crystallization: The driving force 
which causes nucleation corresponds to supersaturation or
supercooling, and  the nucleation energy $\Delta F$ is small for high
supersaturation (supercooling).\cite{crystal}
In this system, the nucleation energy $\Delta F$ is the energy that is
needed to create a negative-$\phi$ (black) domain in the ``sea'' of
positive-$\phi$ spins (white space).  
However, there is no defined quantity called
supersaturation or supercooling. Let us consider a quantity that
expresses the degree of non-equilibrium as ``supersaturation.'' The
supersaturation should depend on the  external field and the field
sweeping rate. 
Since the changing rate of the valuable $\phi(\mib{r})$ is finite, 
the system cannot always keep up with the change in the field.  
It is natural to expect the following: the larger the value of $v$, 
the higher is the supersaturation. 
Therefore, when $v$ is large, $\Delta F$ is small because of high
supersaturation and nucleation easily occurs, which leads to
the sudden appearance of many nuclei. On the contrary, when $v$ is small,
only a small number of nuclei appear because of large $\Delta F$. 
Once the negative-$\phi$ domains appear, the domains grow but new domains
hardly appear. 

The supersaturation as well as the nucleation energy $\Delta F$ should
be related
to the hysteresis loop. In ref.~\citen{jang}, it is shown that the shape
of the hysteresis loop changes depending on the sweeping rate of
the field. The
area of the hysteresis loop is equal to the energy dissipated per
cycle. The area is large for a fast sweep and small for a slow
sweep.\cite{jang} Though the hysteresis and dynamic phase transition
were 
mainly studied for different models in refs.~\citen{luse,sides,jang},
the methods used in the references will be helpful to evaluate 
$\Delta F$. The quantization of $\Delta F$ will be discussed sometime in
the future.
 
There is another question to be considered: how is it that a labyrinth
structure is observed in some samples even when the external field is
suddenly removed? 
This can be explained by the difference in the proper time scale of the
individual samples in the absence of the applied field. The proper time
scale is determined by the anisotropy, interactions between spins, and
disorder in the samples. The difference in the time scale may be
interpreted as the difference in ``the sensitivity to the dynamical
change in the field.''
Using the concept of the \textit{sensitivity} to the field change, we
can consistently understand both our results and the above fact that 
a labyrinth structure often appears under a sudden decay of the field.  
Our results suggest that the \textit{sensitivity}, in the same sample,
should depend on the field sweeping rate $v$.
On the other hand, if a sample has very high \textit{sensitivity} to the
field change, a
labyrinth structure can be observed in the sample even when the external
field is suddenly removed. 
In other words, the balance between the \textit{sensitivity} and the
field sweeping rate is one of the factors that causes a variety of domain
patterns. 

In conclusion, we have confirmed that the characteristics of the magnetic
domain patterns under a zero field depend on the decreasing rate of the
external field in both the experiments and the numerical simulations.
For a rapidly decreasing field, many short domains form a sea-island
structure. On the contrary, a small number of long domains form a
complex labyrinth structure for a slowly decreasing field.
Those properties are explained qualitatively by using the idea of
crystallization.  
In other words, in the case of large $v$, high ``supersaturation''
lowers the nucleation energy, which causes a
sudden appearance of many nuclei, i.e., negative-$\phi$ domains, in a
sea of positive-$\phi$ spins. 
However, a labyrinth structure appears even when the external field
is suddenly removed in some experiments because the spins in actual
garnet films usually vary fast enough to keep up with the change in the
field. 
The results in this letter show that the characteristics of domain
patterns are dependent on the field sweeping rate, or more generally, 
on the balance between the rate and a dynamical property
such as ``the sensitivity to the dynamical change in the field.''

\section*{Acknowledgment}

One of the authors (K. K.) is supported by JSPS Research
Fellowships for Young Scientists.


\begin{thebibliography}{99}
\bibitem{note1} We have observed such behavior in numerical
	simulations. The observations will appear in the future work.
\end{thebibliography}
\end{document}